\documentstyle[12pt]{article}
\begin{document}
\title{
Singularities, initial and boundary problems of the Tolman-Bondi
model}
\author{
Alexander Gromov \\ \\
\small St. Petersburg State Technical University, Faculty of Technical \\
\small Cybernetics, Dept. of Computer Science 29, Polytechnicheskaya str.\\
\small St.-Petersburg, 195251, Russia. e-mail: gromov@natus.stud.pu.ru
}
\date{}
\maketitle
\thispagestyle{empty}
\begin{abstract}
Boundary problem for Tolman-Bondi model is formulated.
One-to-one correspondence between singularities
hypersurfaces and initial conditions of the Tolman-Bondi model is
constructed.
\\ PACS number(s): 95.30.Sf, 98.65.-r, 98.80.-k
\\ keywords: cosmology:theory --- gravitation ---large- scale structure of
universe --- relativity
\end{abstract}
\newpage

\markright{}
\section{Introduction} \label{1}

Two theorems about asymptotic behavior of the Tolman-Bondi model have been
prooved by Olson and Silk (1979) with the assumption of no bang-time
variation:
\begin{equation}
t_0(r) = constant,
\label{const}
\end{equation}
where $r$ is a co-moving coordinate.
A general Tolman-Bondi model has been completely determined by the two
basic functions: $t_0(r)$ and full specific energy $E_0(r)$. The formal
sense of the functions $t_0(r)$ and $E_0(r)$ is the extra conditions of the
Tolman-Bondi model.
In the paper (Olson and Silk)
the equation (\ref{const}) determines the function $t_0(r)$.

In the paper (Gromov 1997) the Tolmn-Bondi model is formulated as the
Cauchy problem for the equation of motion where initial conditions of the
model are initial density $\rho_0(r)$ and energy $E_0(r)$ profiles.

The present paper is dedicated to generalisation of Olson and Silk's
idea (use the bang-time function as extra condition) in the frame of
the approach (Gromov 1997). It is shown that the Tolman-Bondi model admits
two different formal statements: initial and boundary problems. On this
ground one-to-one correspondence
between two singularities of density and two initial conditions of the
Tolman-Bondi model is constructed.

\markright{}
\section{Cauchy problem for the Tolman-Bondi \newline model}
\markright{}

In this section the Tolman-Bondi model (using $c = 1$ and $G = 1$)
is represented in the same way as in the paper (Gromov 1997).
The co-moving coordinate 'invariant mass' $M_i$ is used here. Using
the $M_i$ coordinate the Tolman-Bondi model is represented as the Cauchy
problem for the equation of motion:
\begin{eqnarray}
2\,\ddot R(M_i,t)R(M_i,t) + \dot R^2(M_i,t)
+ 1 - f^2(M_i) = 0, \qquad \dot{} = \frac{\partial}{\partial \tau}
\label{eq-m}
\end{eqnarray}
with initial conditions:
\begin{eqnarray}
R(M_i,0) =
R_{0}(M_i) = \left[
\frac{3}{4\,\pi}\,\int\limits_{0}^{M_i}
\frac{f(x)}{\rho(x,0)}\,{\rm d} x
\right]^{1/3},
\label{Mm1}
\end{eqnarray}
\begin{eqnarray}
\dot R(M_i,0) = \dot R_0(M_i),
\label{velocity}
\end{eqnarray}
where $R(M_i,t)$ is the Euler radial coordinate of the particle;
$\rho_0(M_i,t)$ is the
density; $\dot R(M_i,t)$ is the velocity; $\rho_0(M_i,t)$ and
$\dot R(M_i,t)$ are given functions.

The function  $f(M_i)$ is defined by the equation
\begin{eqnarray}
\frac{1}{2}\,\left(\frac{3}{4\,\pi}\right)^{1/3}\,
\left(\int\limits_{0}^{M_i}
\frac{f(x)\,{\rm d}x}{\rho_0(x)}\right)^{1/3}\,
\left[\dot R^2_0(M_i) - f^2(M_i) + 1\right] &=&
\label{f def1}
\end{eqnarray}
$$\int\limits_{0}^{M_i} f(x){\rm d}x.$$

The general solution of the equation (\ref{eq-m}) in the frame of
approach (Gromov 1997) is:
\begin{eqnarray}
\pm t + t_0(M_i) = \int\limits_{0}^{R(M_i,t)}
\frac{{\rm d} \tilde R}{\sqrt{
f^2(M_i) - 1 + \frac{2}{\tilde R}\,\int\limits_{0}^{M_i}
f(\tilde M_i) {\rm d} \tilde M_i
}}
\label{g s}
\end{eqnarray}
where
\begin{eqnarray}
t_0(M_i) = \int\limits_{0}^{R_0(M_i)}
\frac{{\rm d} \tilde R}{\sqrt{
f^2(M_i) - 1 + \frac{2}{\tilde R}\,\int\limits_{0}^{M_i}
f(\tilde M_i) {\rm d} \tilde M_i
}}.
\label{g s 1}
\end{eqnarray}
Function $t_0(M_i)$ is the bang-time function which has been used by Olson
and Silk (1979).

\markright{}
\section{Boundary problem for the Tolman-Bondi model}
\markright{}

In this section the Tolman-Bondi model is studied as the boundary problem
for the equation of motion.

The interval of the Tolman-Bondi space-time is:
\begin{equation}
{\rm d}s^2(M_i,t) =
{\rm d}t^2 - \frac{R^{\prime\, 2}(M_i,t)}{f^2(M_i)}\,{\rm d}r^2
 - R^{2}(M_i,t)\,{\rm d} \Omega^2,
\label{metric-1}
\end{equation}
where
\begin{equation}
{}^{\prime} = \frac{\partial}{\partial M_i}, \qquad
{\rm d}\Omega = {\rm d}\theta^2 + sin^2\theta {\rm d}\phi,
\label{metric-2}
\end{equation}
so that for metric coefficients we obtaine:
\begin{equation}
g_{11}(M_i,t) = \left( \frac{R^{\prime}(r,t)}{f(r)} \right)^2,
\label{g_11a}
\end{equation}
\begin{equation}
g_{22}(M_i,t) = g_{33}(M_i,t) = R^2(M_i,t).
\label{g_22a}
\end{equation}

The density is given by the equation
\begin{eqnarray}
4 \pi \rho(M_i,t) = \frac{f(M_i)}{R^2(r,t)\,
\displaystyle\frac{\partial R(M_i,t)}{\partial M_i}}.
\label{T:15:1}
\end{eqnarray}
Formula (\ref{T:15:1}) becomes identity on the set of initial conditions
(\ref{Mm1}).
Two singularities of density
\begin{eqnarray}
\displaystyle\frac{\partial R(M_i,t_2)}{\partial M_i} = 0.
\label{s2}
\end{eqnarray}
and
\begin{eqnarray}
R(M_i,t_1) = 0
\label{s1}
\end{eqnarray}
corresponds to two hypersurfaces in the space-time where metric
coefficients are equal to zero:
\begin{equation}
g_{11}(M_i,t) = 0,
\label{g_11}
\end{equation}
\begin{equation}
g_{22}(M_i,t) = g_{33}(M_i,t) = 0
\label{g_22}
\end{equation}
and metric is singular:
\begin{equation}
g = 0.
\label{g_22c}
\end{equation}
Let us name these hypersurfaces 'singular hypersurfaces'.
Following Olson and Silk (1979) let us name also
function the $t_2(M_i)$ from (\ref{s2}) 'second bang-time function'.
Equation (\ref{s1}) gives one of the boundary conditions for the equation
of motion. To transform equation (\ref{s2}) to the form-like (\ref{s1}),
let us use the general solution of the Tolmn-Bondi model
(\ref{g s}).
First we rewite the solution (\ref{g s}) in the form
\begin{eqnarray}
\left[\pm t + t_0(M_i) \right]
\sqrt{\int\limits_{0}^{M_i} f(\tilde M_i){\rm d} \tilde M_i}
=
\int\limits_{0}^{R(M_i,t)}
\displaystyle\frac{{\rm d} \tilde R}{\sqrt{
\displaystyle\frac{f^2(M_i) - 1}{\int\limits_{0}^{M_i} f(\tilde M_i) {\rm d} \tilde M_i
} + \displaystyle\frac{2}{\tilde R}
}}
\label{g s new}
\end{eqnarray}
This equation defines the function $R(M_i,t)$ implicitly. After
diffirentiation $\frac{\partial}{\partial M_i}$ and using equality
(\ref{s2}) we obtain the following equation for $R(M_i,t_2)$:
\begin{eqnarray}
\pm \frac{t_2\,R(M_i,t_2)}{
2\,\sqrt{\int\limits_{0}^{M_i} f{\rm d}\tilde M_i}
}
+
\frac{R_0^{\prime}(M_i,t_2)}{
\sqrt{
\frac{f^2(M_i) - 1}{\int\limits_{0}^{M_i}f(\tilde M_i) {\rm d}\tilde M_i}
+ \frac{2}{\tilde R_0}
}
}
= \nonumber\\
\frac{R^{\prime}(M_i,t_2)}{
\sqrt{
\frac{f^2(M_i) - 1}{\int\limits_{0}^{M_i}f(\tilde M_i) {\rm d}\tilde M_i}
+ \frac{2}{\tilde R(M_i.t_2)}
}
} = 0
\label{44}
\end{eqnarray}
at time $t = t_2$,
where $t_2$ is time when equation (\ref{s2}) is satisfied.

Let us study the flat TB model: $f(M_i) = 1$. Equation (\ref{s1})
corresponds to
\begin{eqnarray}
t_1(M_i) = t_0(M_i)
\label{45a}
\end{eqnarray}
and (\ref{s2}) corresponds to
\begin{eqnarray}
t_2(M_i) = \sqrt{2\,M_i\,R_0(M_i)}\,R^{\prime}_0(M_i),
\label{45}
\end{eqnarray}
that coinsides with the results of paper (Gromov 1997).

$R(M_i,t_1) = 0$ and $R(M_i,t_2)$ from (\ref{44}) have now the sense of
boundary conditions for the equation of motion.
\newpage
\markright{}
\section{Results and discussion}
\markright{}

The TB model is reduced to the ordinary differential equation of motion
(Gromov 1997) with extra conditions (initial or boundary) and completely
defined by two functions of the set
\begin{eqnarray}
t_1(M_i), t_2(M_i), \rho_0(M_i), f(M_i), R_0(M_i), \dot R_0(M_i),
\label{set}
\end{eqnarray}
but not every two of them can be used simultaneously. Every two functions
from the set
\begin{eqnarray}
\rho_0(M_i), f(M_i), R_0(M_i), \dot R_0(M_i)
\label{subset1}
\end{eqnarray}
produce Cauchy problem, but two bang-time functions
\begin{eqnarray}
t_1(M_i), t_2(M_i)
\label{subset2}
\end{eqnarray}
produce boundary problem for the equation of motion.

In accordance with the theorem about a correspondence between intial and
boundary problems (Korn G.A. and Korn T.M, 1968 ) we can transform bondary
problem into initial
one, or intial problem into boundary one. This is supported by the formulas
(\ref{44}) and (\ref{g s 1}): if $t_1(M_i)$ and $t_2(M_i)$ are specified,
we can calculate $\rho_0(M_i)$ and $f(M_i)$ and leave initial from boundary
problem to the initial one.
So,
{\it
it is not possible to neglect one of
two singularities of the TB model because it breaks the formal statement of
the problem.
}
We see also that two bang-time functions define two singular hypersurfaces
of the TB model.
Formulas (\ref{44}) and (\ref{g s 1})
produce one-to-one correspondence between two set of functions
(\ref{subset1}) and (\ref{subset2}).
So, from the theorem it follows the
\newline {\it implication:

One-to-one correspondence between singularities hypersurfaces and initial
conditions of the Tolman-Bondi model exists.
}

The Tolman-Bondi model has spherical symmetry. This leads to the equality
\begin{eqnarray}
g_{22}(M_i,t) = g_{33}(M_i,t),
\label{g 22 33}
\end{eqnarray}
which corresponds to singularity $R(M_i,t_1) = 0$.
One of the possible generalisations of the Tolman-Bondi model is small
perturbation of the spherical symmetry, that provides the break down of the
conditions (\ref{g 22 33}):
\begin{eqnarray}
g_{22}(M_i,t) \ne g_{33}(M_i,t),
\label{g 22 NE 33}
\end{eqnarray}
so one singularity $R(M_i,t_1) = 0$ will be split by the two different
singularities like this:
\begin{eqnarray}
g_{22}(M_i,t) = 0,
\label{g 22 NE}
\end{eqnarray}
and
\begin{eqnarray}
g_{33}(M_i,t) = 0.
\label{g NE 33}
\end{eqnarray}

\markright{}
\section{Acknowledgements}
\markright{}

I am grateful for encouragement and discussion
to Prof. Arthur Chernin,  Prof. John Moffat,  Dr. Yurij Baryshev
and Marina Vasil'eva.
\newline
This paper was financially supported by "COSMION" Ltd., Moscow.
\newline

{\bf REFERENCES}
\newline

Bondi H. (1947) MNRAS, {\bf 107}, 410.

Gromov A. (1997)  gr-qc/9612038;

Olson D. and Silk J. (1979) AphJ, {\bf 233}, 395.

Tolman R.C. (1934) Proc.Nat.Acad.Sci (Wash), {\bf 20}, 169.

Korn G. and Korn T. (1968) Mathematical Handbook.

\end{document}